\documentclass[a4paper,11pt]{article}
\pdfoutput=1 

\usepackage{jinstpub} 
\usepackage{hyperref}
\usepackage{caption}
\usepackage{subcaption}

\title{\boldmath Light detection with power and signal transmission over fiber}

\author[a,1]{H. V. Souza\note{Corresponding author.}}

\affiliation[a]{Laboratoire Astroparticule et Cosmologie,\\Paris, France}

\emailAdd{hvsouza@apc.in2p3.fr}

\abstract{The Deep Underground Neutrino Experiment (DUNE) is a next generation long baseline (1300~km) neutrino oscillation experiment. The neutrino beam measurements will be performed by a near detector (ND) and far detector (FD). The far detector will consist of four modules, installed 1,5 km deep underground, based on Liquid Argon Time Projection Chamber (LArTPC) technology to detect particles.

The Vertical Drift (VD) LArTPC is a recent technology proposed by the DUNE Collaboration for the second FD module. In VD, light collection will be optimized by embedding photon detectors within the LArTPC cathode, which is biased at --300~kV. As result, power must arrive to the Photon Detection System (PDS) and signal must be transmitted via non-conductive material. The proposed solution is to use Power-over-Fiber (PoF) and Signal-over-Fiber (SoF). An intense validation of the system is being performed by the Collaboration at the CERN Neutrino Platform, design and results from data collected over the first half of 2022 are presented. }


\arxivnumber{1234.56789} 

\collaboration[c]{on behalf of the DUNE collaboration}

\proceeding{LIDINE2022 - Light Detection In Noble Elements\\
  21-23 September 2022 \\
  University of Warsaw Library}

\begin{document}
\maketitle
\flushbottom

\section{Introduction}
\label{sec:intro}

The Deep Underground Neutrino Experiment~(DUNE)~\cite{DUNE_all_TDR} is a long baseline neutrino experiment with the primary goal of performing precise measurements of the neutrino oscillation parameters, determining the neutrino mass hierarchy and the Charge Parity (CP) violating phase in the lepton sector. The detector has also capability to perform excellent supernova neutrino burst detection and to investigate proton decay. DUNE consists of a near and a far detector, the Near Detector (ND)~\cite{near_detector} will be placed around 574~m from the neutrino beam generated at the Long-Baseline Neutrino Facility~(LBNF) at the Fermi National Accelerator Laboratory (Fermilab). The Far Detector~(FD) is located at the Sanford Underground Research Facility (SURF) 1300~km away and around 1.5~km underground. The FD is divided in four modules of Liquid Argon Time Projection Chambers~(LArTPC)~\cite{LAr_fund_properties} each with 17.5~kt in mass. 

\begin{figure}[htbp]
\centering
\includegraphics[width=0.8\linewidth]{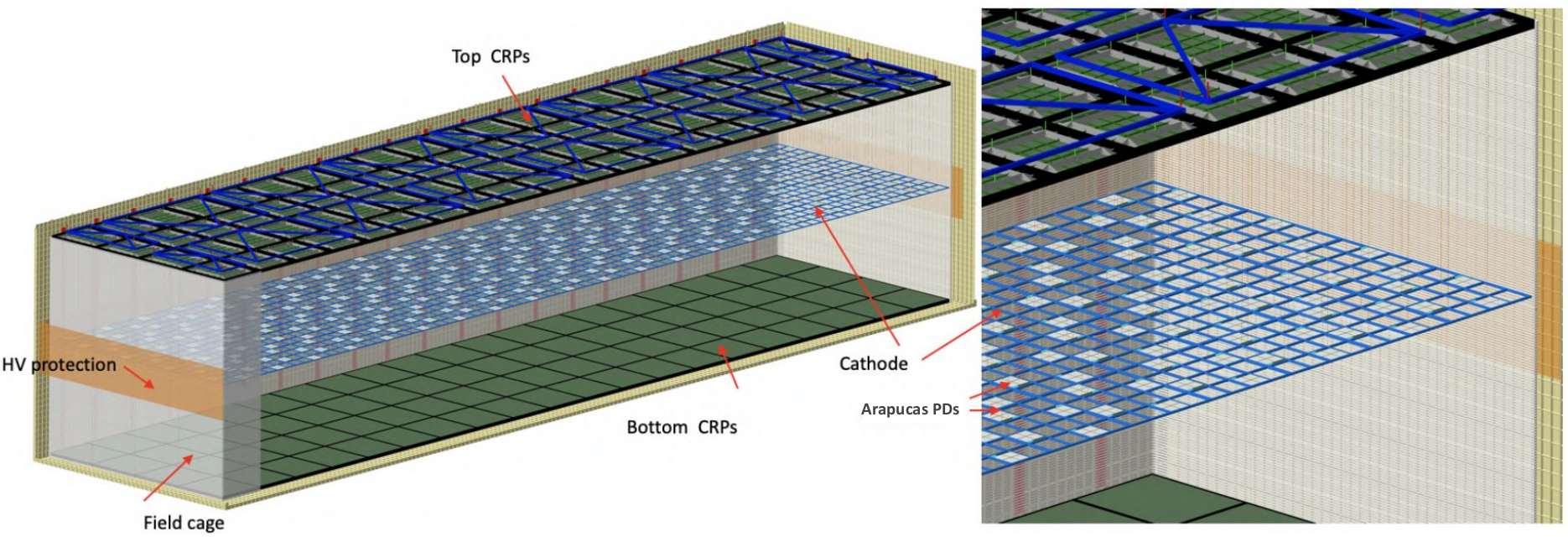}
\caption{\label{fig:VD_TPC_schematic}Schematic concept of the Vertical Drift module~\cite{VD_CDR}. (Left) The Charge Readout Planes~(CRP)~(Anode) are at the top and bottom of the LArTPC with a 6.5~m distance between the cathode. The Field Cage ensures uniform Electric Field. (Right) Photon Detection system, based on the X-Arapuca device~\cite{propostaARA,x_arapuca,x_arapuca_article}, is installed at the cathode.}
\end{figure}

A LArTPC based on the Vertical Drift~(VD) technology~\cite{VD_CDR} is being developed by the collaboration since 2020 as the second FD module. In this design, electrons are drifted vertically towards the anodes on the top and bottom of the detector, while the cathode is placed horizontally in the middle. The drift distance of 6.5~m is achieved by biasing the cathode to --300~kV. Figure~\ref{fig:VD_TPC_schematic}
 shows the schematic of the 62$\times$15.1$\times$14~m$^3$ LArTPC, where the anodes consist of Charge-Readout Planes~(CRP), formed by perforated printed circuit boards~(PCB), to collect the drifting electrons. The photon detectors are X-Arapuca device~\cite{propostaARA,x_arapuca,x_arapuca_article,VD_PDS_paulucci_2022} and are placed inside the cathode. The power and analog readout of the Photon Detection System (PDS) must be done through non-conductive materials as the latter are in common with the --300~kV bias. This work focus in the Power and Signal over Fiber, the proposed solution for the VD light detection.

\section{Photon detectors} 

\begin{figure}[htbp]
	\centering
	\begin{subfigure}{0.45\textwidth}
	    \centering
	    \includegraphics[width=0.99\linewidth]{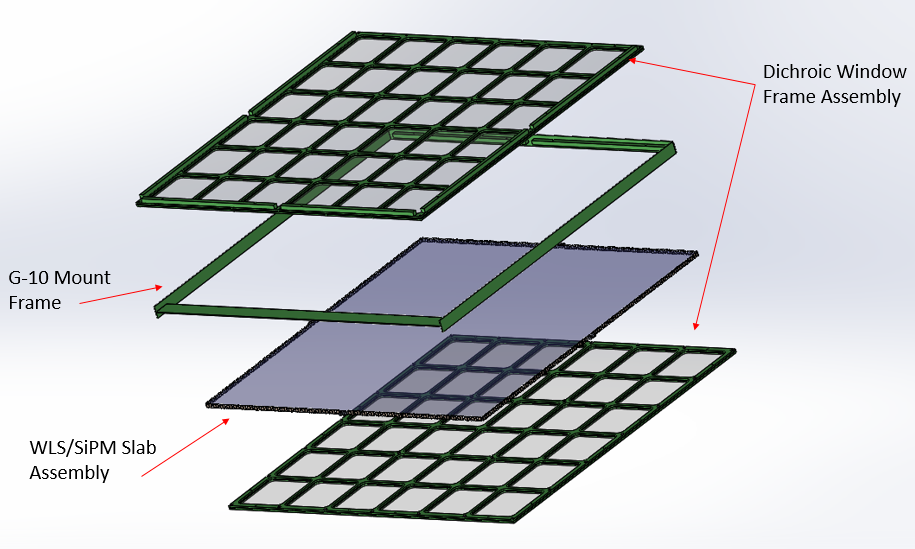}
	    \caption{\label{fig:xarapucacell}}
	\end{subfigure}
	\begin{subfigure}{0.45\textwidth}
	    \centerline{\includegraphics[width=0.99\linewidth]{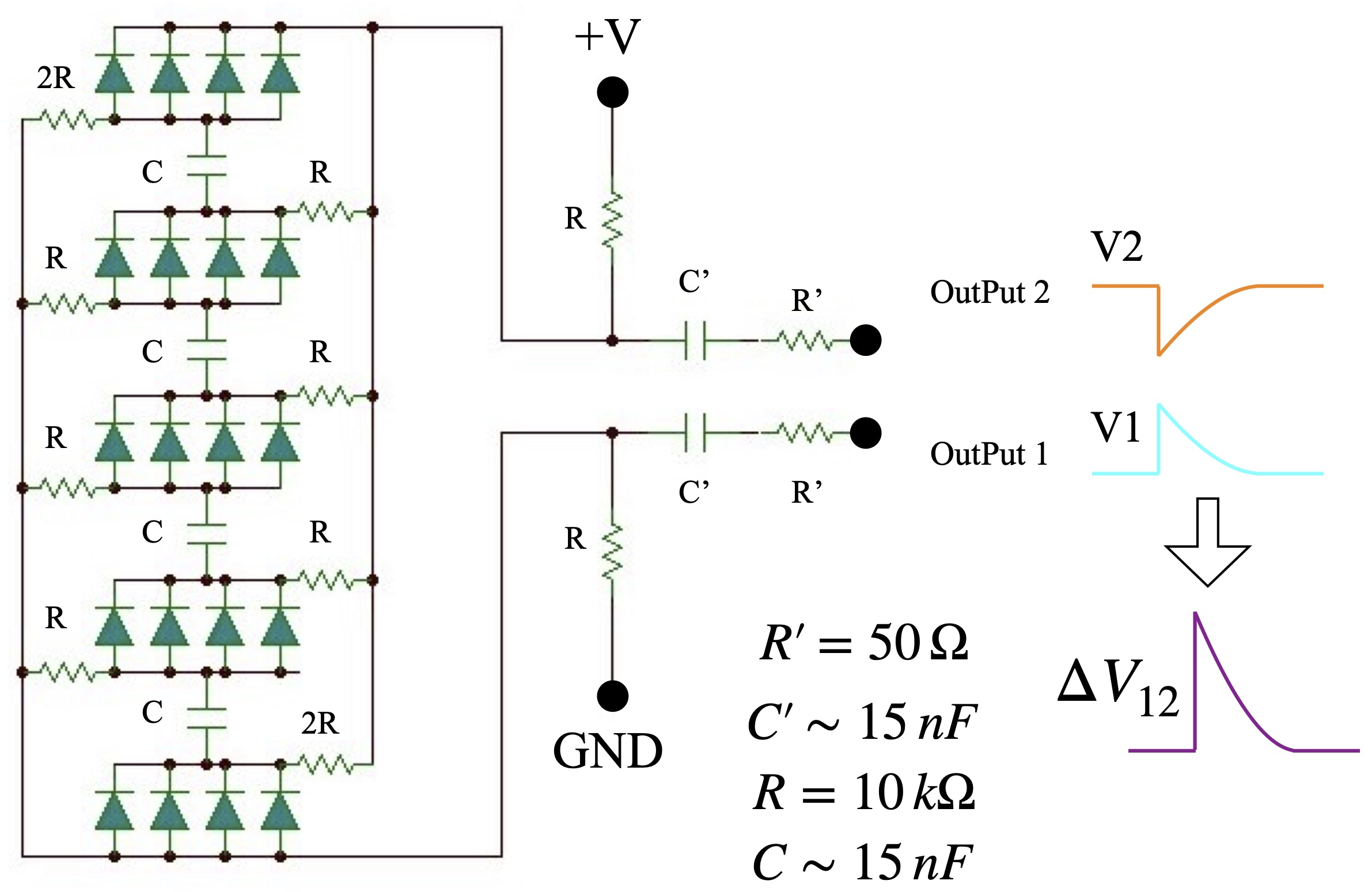}}
	    \caption{\label{fig:sipms_gang}}
    \end{subfigure}
    \caption{a) Schematic of the X-Arapuca tile. The wavelength shifter plate is surrounded by SiPMs and placed in between 2$\times$36 coated dichroic filters. b) Passive hybrid ganging scheme of the SiPM board. Bias voltage (DC) and signals (AC) are decoupled~\cite{wolteHenrique}.}
\end{figure}

The Photon Detection System (PDS) of the DUNE's second FD module will be based on the X-Arapuca technology~\cite{propostaARA,x_arapuca,x_arapuca_article}: a light trapping device which uses a combination of wavelength shifters (WLS) and dichroic filter to trap photons inside a highly reflective cavity to be detected by silicon photomultipliers~(SiPM). The schematic of a X-Arapuca tile is shown in Figure~\ref{fig:xarapucacell}, the device sizes 60$\times$60~cm$^2$ and 2.5~cm thick where two sets of 36 dichroic filters coated with an external WLS are placed on the top and bottom of an internal WLS, which makes the final conversion of the photons wavelength and guide them to 160 SiPMs, nested in the sides of the WLS plate.

The SiPMs are divided in two channels, with 80 SiPMs each, and the small group of 20 SiPMs are passively ganged in a hybrid configuration for biasing and signal readout, in parallel and series respectively, as shown in Figure~\ref{fig:sipms_gang}. The differential signal from anode and cathode are AC-coupled to the readout. 

\section{Cold electronics board}
\subsection{Signal over Fiber (SoF)}
\label{sec:sof}

The analog signal transmission over fiber is the proposed solution to readout the light signals from the SiPMs through a non-conductive material. The technology is commonly applied for warm experiments and industry. However no commercial solution exists that can function within cryogenic liquids so a custom-make cold electronics (CE) board has been developed~\cite{Sacerdoti_2022}.

\begin{figure}[htbp]
    \centering
    \includegraphics[width=0.8\linewidth]{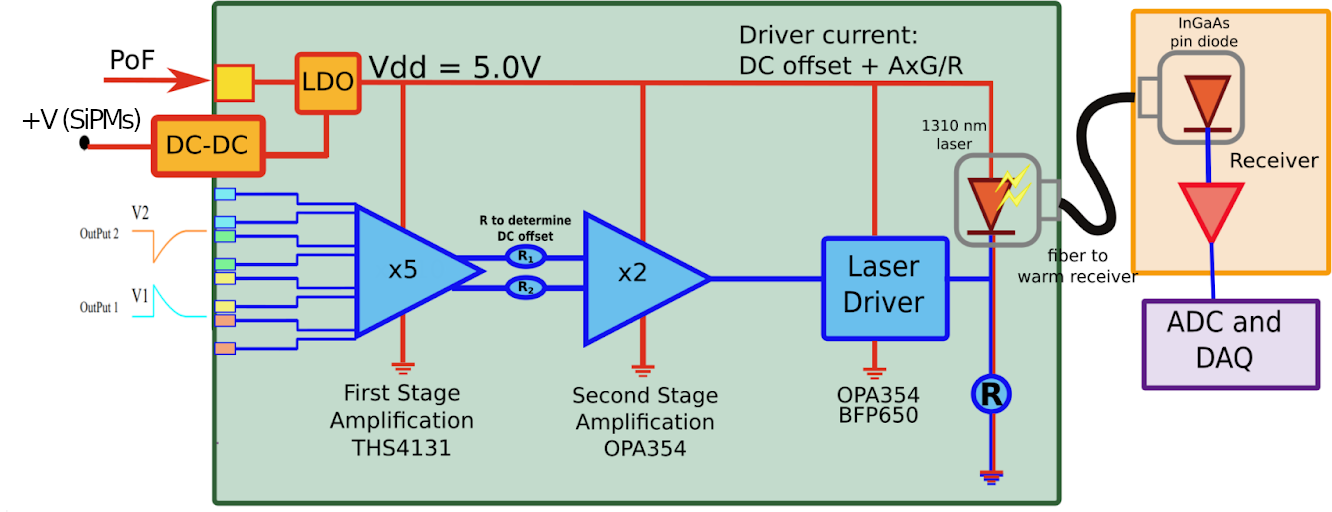}
    \caption{Schematic of the CE board with the SiPM bias through DC-DC converters and SoF.}
    \label{fig:board_schematics}
\end{figure}

The latest schematic version of the CE board is presented in Figure~\ref{fig:board_schematics}. The electric power is given by the Power over Fiber~(see Section~\ref{sec:pof}) to a low-dropout (LDO) regulator which outputs a stable $\sim$5.2V for the active components of the board and to the DC-DC converter. The DC-DC converts the low voltage to a high voltage and low current for the biasing of the SiPMs.

The differential signal from the SiPMs goes through the first stage amplification with a gain $\sim$5 and is converted to an unipolar signal at the second amplification stage with gain $\sim$2 (resulting in a total again about 20 times). The signal is passed to the third stage, a laser driver which controls the 1310~nm Fabry-Pérot diode laser current. The laser is kept with a constant offset, in order to be operated in the linear region. A set of resistors in parallel, with the use of a NTC thermistor allows the use of the CE board in cold and room temperature. The analog signal is transmitted through optical fibers to the warm receivers (Koheron PD100 units) to be converted in electrical signals again and readout by the DUNE data acquisition (DAQ).

The PDS system has the requirement to operate with a Signal-to-Noise ratio (SNR) $>$4, a bandwidth of 30~MHz and a dynamic range of 1000~photo-electrons (p.e.), that is, to be able to detect up to 1000~photons. To accomplish these requirements, the Operational Amplifiers (OpAmps) were selected due to their high bandwidth ($>$150~MHz), low noise, being rail-to-rail and good performance in cryogenic temperatures~\cite{Sacerdoti_2022,wolteHenrique}. The gain of each OpAmp was set to compromise between the SNR and the bandwidth. 

Currently, it is under development: a different amplification stage, with higher dynamic range and lower noise, by setting the third OpAmp as the laser driver without the need of the transistor; improvement in the laser-fiber interface to avoid loss when flooded by liquid argon; a custom made warm receiver which aims to integrate better with the front-end readout electronics of DUNE's first FD module. 

\subsection{Power over Fiber (PoF)}
\label{sec:pof}

The power supply of the SiPMs and the active components is given through fiber. The PoF system was completely developed at Fermilab. The current setup~\cite{wolteHenrique}, uses Gallium Arsenide (GaAs) Photovoltaic Power Converters (PPC). Three PPC receivers are mounted in the CE board in parallel, each receiver can output up to 7~V and a maximum current of $\sim$80~mA. This configuration allows a higher current output for the active components and the DC-DC converter, which is around 100~mA. The GaAs was chosen due to the high ($>$50\%) efficiency conversion even in liquid argon temperature (87.3~K).  The light input is given by a Broadcom module with a +2~W capability 808~nm laser diode through standard multi-mode fiber link (62.5/125 $\mu$m). 

\section{System validation}
\label{sec:sys_val}

\begin{figure}[htbp]
	\centerline{\includegraphics[width=0.466\linewidth]{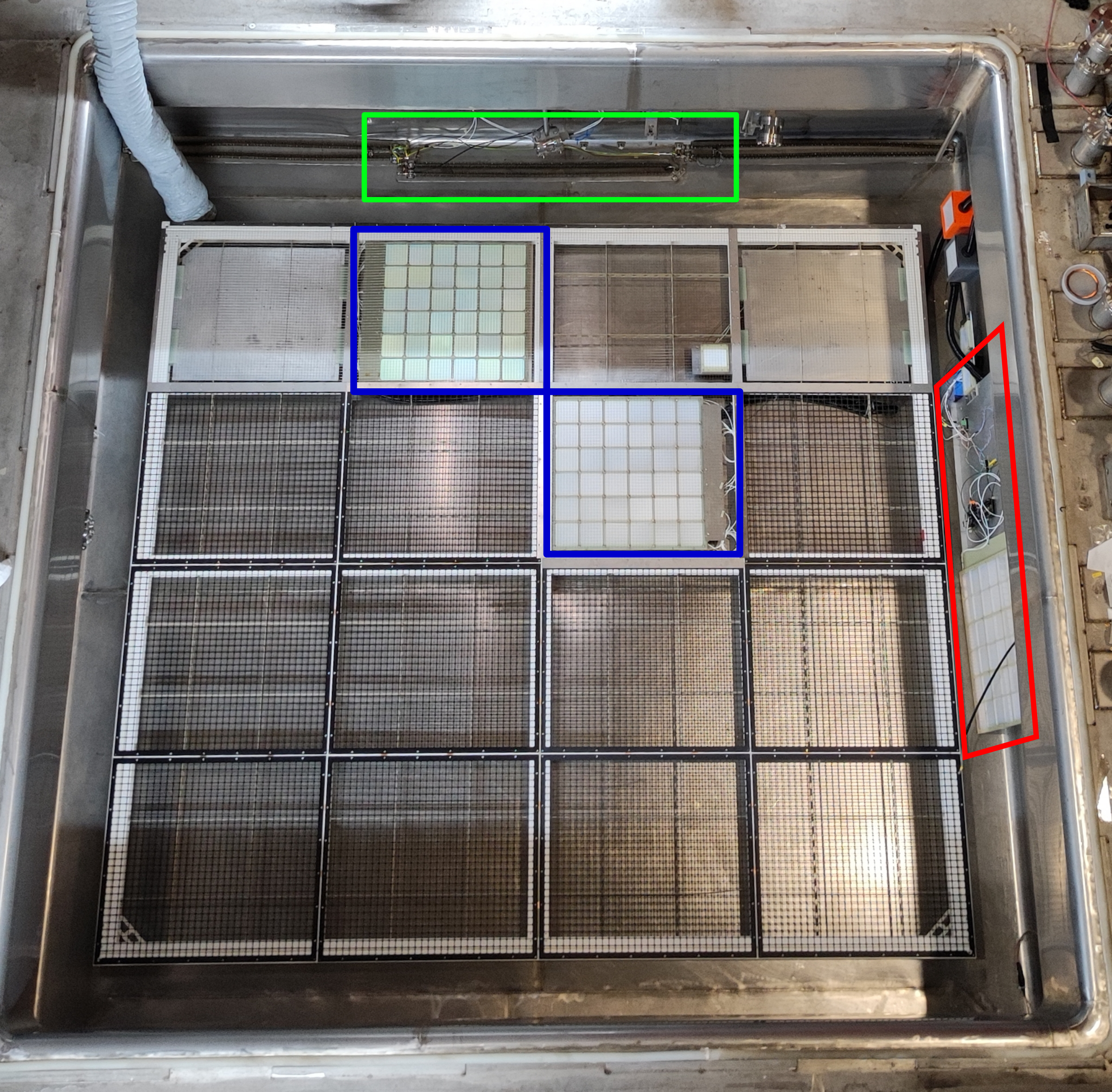}
	\includegraphics[width=0.514\linewidth]{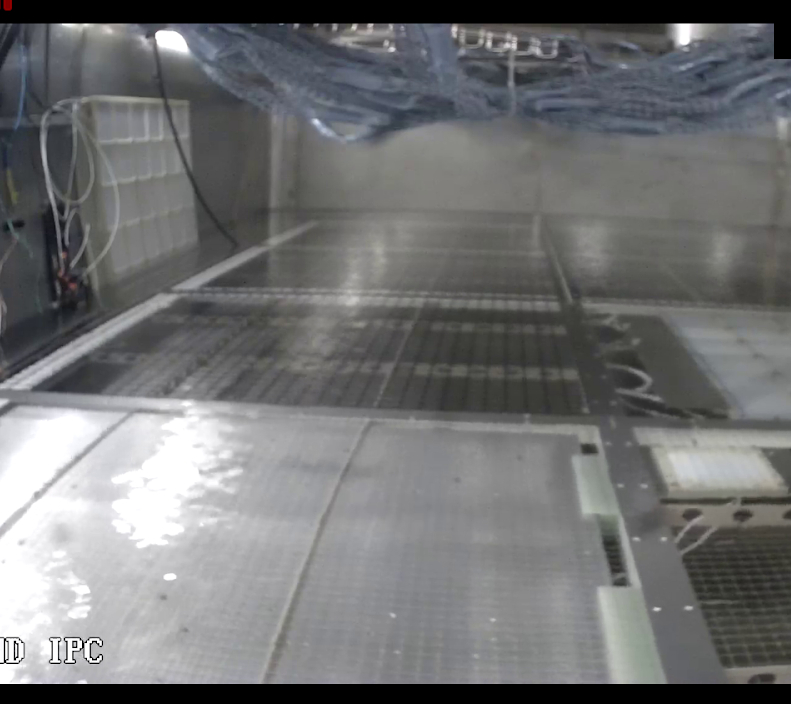}}
	\caption{(Left) Cathode installed in the coldbox with two x-Arapucas (blue squares) using PoF and SoF, LED fiber output for calibration and photon detectors in the membrane for characterization of the system. (Right) Photo of the coldbox half filled with liquid argon, in this photo the CRP is not present.}
	\label{fig:coldbox_cathode_pds_photo}
\end{figure}

The validation and integration of the system in liquid argon is done in the prototype called ``Coldbox'', installed at the CERN Neutrino Platform. The LArTPC prototype is a (3$\times$3$\times$1)~m$^3$ cryostat, 23~cm drift distance between the anode and the cathode. Figure~\ref{fig:coldbox_cathode_pds_photo} shows the coldbox's cathode on the floor, the blue squares are two x-Arapuca devices installed on the cathode, the green square are light diffusers for an UV LED flasher installed for calibration of the PDS and the red square are photon detectors installed on the membrane which are installed to study the performance of the detectors and development of components. 

The x-Arapucas nested in the cathode were intalled with PoF and SoF. The system of the membrane consist of two small (12$\times$12~cm$^2$) Arapuca devices and one X-Arapuca device, the power of the SiPMs is given through copper wires but the signal is readout through SoF. 

The data acquistion (DAQ) is done with CAEN Digitizer DT5730SB (14~bits, 2~V$_\text{pp}$ and 500 MSamples/s). The data were taken in three different sets: (1) UV LED as external trigger for calibration of the SiPMs to retrieve the number of photo-electrons (p.e.) detected, (2) self-trigger at different levels of signal (covering from 30~p.e. up to 100~p.e.) and (3) coincidence with a cosmic muon paddle telescope (CRT). Besides these three set, long (1~millisecond) waveforms are also taken with a random trigger for the study of the frequency response of the system.

\begin{figure}[htbp]
	\centerline{\includegraphics[width=0.9\linewidth]{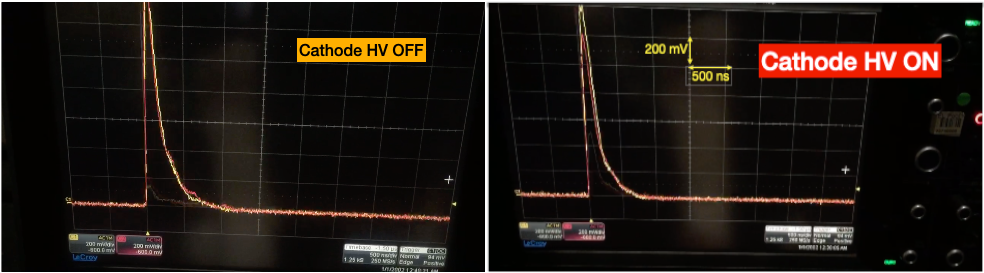}}
	\caption{Readout signals from the x-arapuca operated with PoF and SoF without high voltage (left) and with a high voltage of --10~kV (right)~\cite{wolteHenrique}.}
	\label{fig:pds_signals}
\end{figure}

In December 2021, the first signal transmission in liquid argon of a x-Arapuca device through fiber only was accomplished~\cite{wolteHenrique}. The device was operated with both cathode off and at --10~kV, both with a stable performance as shown in Figure~\ref{fig:pds_signals}.

\begin{figure}[htbp]
	\centering
	\begin{subfigure}{0.505\textwidth}
	    \centering
	    \includegraphics[width=0.99\linewidth]{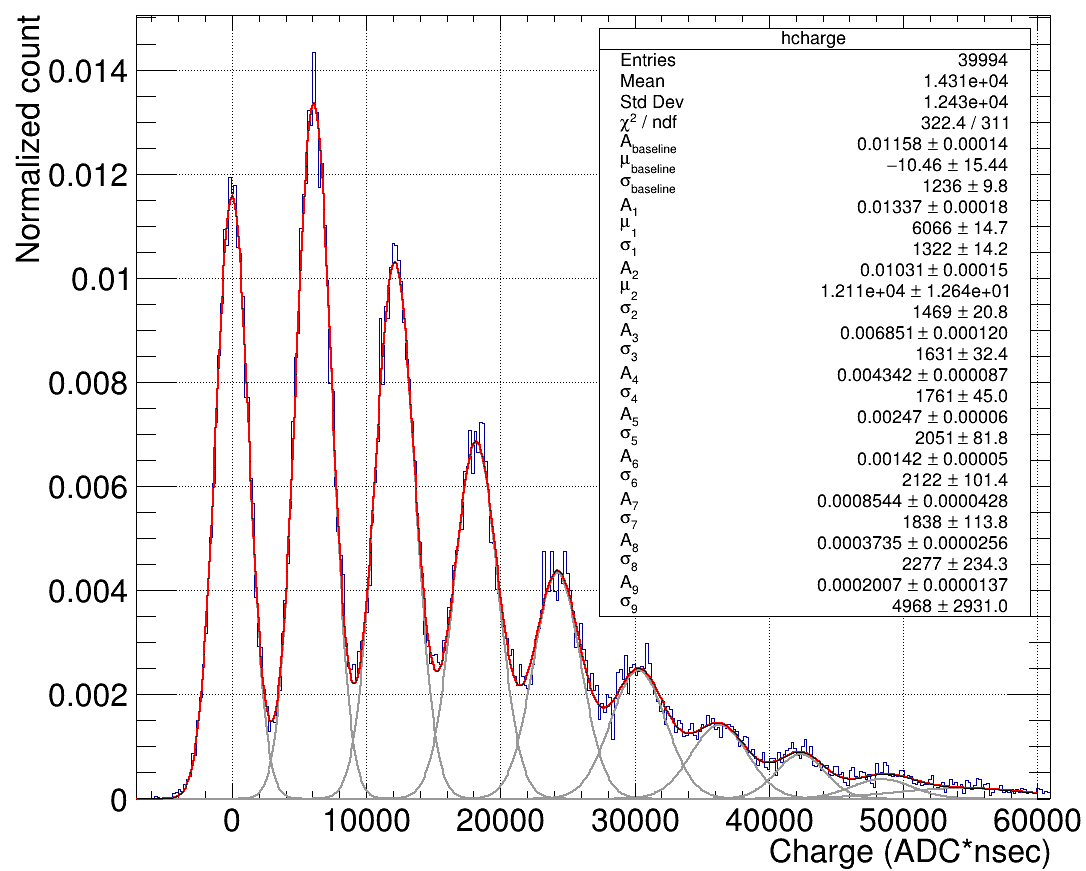}
	    \caption{\label{fig:spe_miniArapuca}}
	\end{subfigure}
	\begin{subfigure}{0.47\textwidth}
	    \centerline{\includegraphics[width=0.99\linewidth]{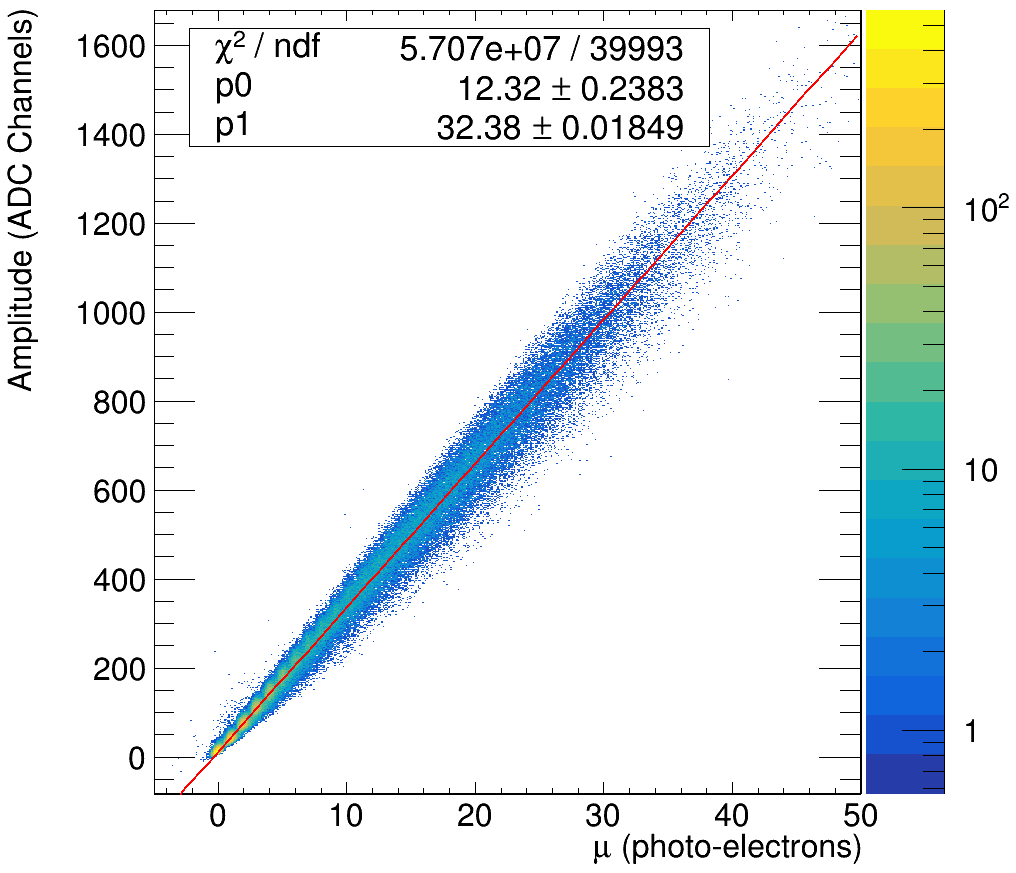}}
	    \caption{\label{fig:miniArapuca_linearity}}
    \end{subfigure}
    \caption{a) Single photo-electron spectrum obtained by flashing an UV LED with low intensity, the signals were readout through SoF and the SiPMs were biased by copper. b) Linear response of the device as amplitude of the signals versus photo-electrons detected.}
\end{figure}

The system was calibrated with the LED pulses, as shown in Figure~\ref{fig:spe_miniArapuca} where the spectrum of photo-electrons was retrieved with a Signal-to-Noise ratio of 4.9\footnote{Signal-to-noise ratio defined as the average charge of one photo-electron divided by the standard deviation of the baseline charge (first peak of the histogram).}. The fit is performed with N+1 Gaussian's, with a baseline (no photo-electrons detected) and N peaks corresponding to N photo-electrons. The amplitude and standard deviation is left as a free parameter for all Gaussians, however the mean $\mu_N$ of N$\ge$3 Gaussians is kept fixed from the distance $\mu_2 - \mu_1$. The linear response of the detector was also studied and it is shown in Figure~\ref{fig:miniArapuca_linearity} which shows the amplitute of the signals, in ADC Channels, versus the average number of photo-electrons detected. The linear response of the device was proved up to $\sim$300 p.e., in which the LED intensity was maximum~\cite{wolteHenrique}.

The PoF and SoF could be improved due to the tests at the Coldbox. Besides the linearity study shown in this work, other measurements and analysis were done with the data, such as: to prove the effect of the electric field in the light yield of liquid argon~\cite{electron_recomb_icarus}, verify the purity of the liquid argon through the triplet state component of scintillation, study the light leakage of the infrared light in the system and study of noise level on the PDS system and CRP.

Multiple runs have been performed at the coldbox since December 20201, with several new developments and improvements. The proof of operation of the VD LArTPC photon detection system with power and signal over fiber has been proved with a high voltage of --10~kV through several tests. 

\section{Conclusion}
For the first time, SiPMs based photon detectors (x-Arapucas) powered using PoF and read-out with SoF has been operated at the CERN Neutrino Platform. In this work, we presented the performance of the SoF which resulted in a SNR$>$4 which allows a precise calibration of the system. There are several ongoing research and development for all the instances of the vertical drift technology and the PDS has shown improvements since the beginning of the tests in 2021. The prototype of the second FD module is foreseen to be installed in the beginning of 2023 with many benefits from the Coldbox tests.

\newpage

\bibliography{bibliography_henrique}

\providecommand{\href}[2]{#2}\begingroup\raggedright\begin{thebibliography}{10}

\bibitem{DUNE_all_TDR}
B.~Abi, R.~Acciarri, M.A.~Acero, G.~Adamov, D.~Adams, M.~Adinolfi et~al.,
  \emph{Deep Underground Neutrino Experiment (DUNE), Far Detector Technical
  Design Report, Volume I - IV},  2020.

\bibitem{near_detector}
A.A.~Abud, B.~Abi, R.~Acciarri, M.A.~Acero, G.~Adamov, D.~Adams et~al.,
  \emph{Deep Underground Neutrino Experiment (DUNE) Near Detector Conceptual
  Design Report},
  \href{https://doi.org/10.3390/instruments5040031}{\emph{Instruments}
  {\bfseries 5} (2021) }.

\bibitem{LAr_fund_properties}
T.~Doke, \emph{{Fundamental Properties of Liquid Argon, Krypton and Xenon as
  Radiation Detector Media}}, {\emph{Portugal. Phys.} {\bfseries 12} (1981) 9}.

\bibitem{VD_CDR}
{The DUNE Collaboration}, \emph{Far Detector 2 - Conceptual Design Report},
  {\emph{in preparation} (2022) }.

\bibitem{propostaARA}
A.~Machado and E.~Segreto, \emph{ARAPUCA a new device for liquid argon
  scintillation light detection}, {\emph{Journal of Instrumentation} {\bfseries
  11} (2016) C02004}.

\bibitem{x_arapuca}
A.~Machado, E.~Segreto, D.~Warner, A.~Fauth, B.~Gelli, R.~M{\'{a}}ximo et~al.,
  \emph{The X-{ARAPUCA}: an improvement of the {ARAPUCA} device},
  \href{https://doi.org/10.1088/1748-0221/13/04/c04026}{\emph{J. Instrum.}
  {\bfseries 13} (2018) C04026}.

\bibitem{x_arapuca_article}
H.V.~Souza, E.~Segreto, A.~Machado, R.~Sarmento, M.~Bazetto, L.~Paulucci
  et~al., \emph{Liquid argon characterization of the X-{ARAPUCA} with alpha
  particles, gamma rays and cosmic muons},
  \href{https://doi.org/10.1088/1748-0221/16/11/p11002}{\emph{Journal of
  Instrumentation} {\bfseries 16} (2021) P11002}.

\bibitem{VD_PDS_paulucci_2022}
L.~Paulucci, \emph{The {DUNE} vertical drift photon detection system},
  \href{https://doi.org/10.1088/1748-0221/17/01/c01067}{\emph{Journal of
  Instrumentation} {\bfseries 17} (2022) C01067}.

\bibitem{wolteHenrique}
H.~Vieira~de Souza, \emph{A Photon Detection System with Power and Signal over
  Fiber},  in \emph{2022 IEEE 15th Workshop on Low Temperature Electronics
  (WOLTE)}, pp.~1--4, 2022,
  \href{https://doi.org/10.1109/WOLTE55422.2022.9882876}{DOI}.

\bibitem{Sacerdoti_2022}
S.~Sacerdoti, \emph{Development of analog signal transmission in {LAr} for
  {DUNE}}, \href{https://doi.org/10.1088/1748-0221/17/01/c01069}{\emph{Journal
  of Instrumentation} {\bfseries 17} (2022) C01069}.

\bibitem{electron_recomb_icarus}
S.~Amoruso, M.~Antonello, P.~Aprili, F.~Arneodo, A.~Badertscher,
  B.~Baiboussinov et~al., \emph{Study of electron recombination in liquid argon
  with the ICARUS TPC},
  \href{https://doi.org/https://doi.org/10.1016/j.nima.2003.11.423}{\emph{Nuclear
  Instruments and Methods in Physics Research Section A: Accelerators,
  Spectrometers, Detectors and Associated Equipment} {\bfseries 523} (2004)
  275}.

\end{thebibliography}\endgroup
\bibliographystyle{JHEP}

\end{document}